\documentclass[12pt]{article}
\input{psfig}

\def\nid{\noindent}
\def\cen{\centerline}
\def\mvs{\vskip 0.2in}
\def\svs{\vskip 0.1in}
\def\s{\phantom{0}}

\begin{document}

\mvs
\cen{\large {\bf THE QUEST FOR THE GOLDEN LENS}}
\svs
\cen{\bf Measurement of the Hubble Constant Via Gravitational Lensing---}
\cen{\bf A Review of the Jodrell Bank ``Golden Lenses'' Workshop}

\mvs
\cen {\it {Liliya L. R. Williams (Institute of Astronomy, Cambridge, UK)}}
\cen {and}
\cen {\it {Paul L. Schechter (Massachusetts Institute of Technology, 
Cambridge, MA, USA)}}

\mvs
\begin{abstract}

Gravitational lensing is now widely and successfully used to study
a range of astronomical phenomena, from individual objects, like 
galaxies and clusters, to the mass distribution on various scales, 
to the overall geometry of the Universe. Here we describe and assess 
the use of gravitational lensing as ``gold standards'' in addressing
one of the fundamental problems in astronomy, the determination
of the absolute distance scale to extragalactic objects. This is 
commonly parameterized by the Hubble constant, $H_0$, the current 
expansion rate of the Universe.  The elegance of the underlying 
geometrical principle of the gravitational lensing method combined 
with the recent advances in observations and modelling makes it a 
very promising technique for measuring $H_0$.
\end{abstract}

\mvs
\section{Introduction}

The idea that the images of distant objects might be distorted and
even multiply imaged by the gravitational potentials of intervening
objects is one of the earliest predictions of general relativity.
Gravitational lenses, perhaps better described as gravitational mirages,
are now proving powerful tools in cosmological studies.

The equations of general relativity can be cast in a form where the
gravitational potential acts like an index of refraction.  As in the
case of terrestrial mirages, variations in this index of refraction
cause distortions of bundles of light rays, and occasionally permit
viewers to see several such bundles which were emitted in different 
directions. The result is a multiply imaged (or strongly lensed)
gravitational lens system.

Perhaps 50 distant active and normal galaxies are known, at present,
to be strongly lensed by mass concentrations lying roughly along what 
would otherwise be the shortest path between the galaxy and the Earth.  
The most spectacular case, shown in figure 1, is that of the
cluster of galaxies CL0024+1654 which produces five readily 
recognized images of a more distant ring galaxy.  In this image the
background lensed object is identified by its blue colour, while 
the cluster member galaxies appear reddish yellow. The cluster distorts 
each of the five images somewhat differently, but these distortions are 
easily understood as the effect of the cluster's gravitational potential.

One important use of such systems is to map the gravitating mass in
the lensing object.  Another use is to magnify features in distant
objects which would otherwise be too small to be resolved. Several
star forming galaxies at very high redshift have been studied in 
unprecedented detail thanks to the resolution of these
megaparsec-scale natural telescopes.

In addition to studying individual objects, be they sources or lenses,
multiply imaged systems can  be used to derive global cosmological 
parameters: the matter density of the Universe, $\Omega$; the 
cosmological constant, or the vacuum contribution to the energy content 
of the Universe, $\Lambda$; and the expansion rate of the Universe, 
$H_0$. The first two of these dictate the overall geometry of the 
Universe, i.e. how distances to objects relate to their measured 
redshifts. Because lenses and sources populate a wide range of redshifts,
statistical properties of large samples of lensed objects, such as the 
distribution of their source and lens redshifts and angular separation of
images, depend on the geometry of the Universe. For example, a non-zero 
cosmological constant increases the volume of space at high redshifts, 
hence the abundance of distant lensed systems can be used to estimate 
$\Lambda$. 

While statistics of complete samples of lensed objects are needed to 
measure $\Omega$ and $\Lambda$, a single lensed system, such as a 
multiply-split high redshift QSO (quasi-stellar object, or quasar) 
can in principle yield the value of $H_0$. 
In such a situation the light travel times for the bundles of light 
from multiple images are different, permitting the observer to see the 
emitting source at two different moments in time. The resultant time 
difference, or time delay, depends upon the differences in the path 
lengths among the multiple light paths.  Because all of the angles are 
known in many lensed systems, a difference in path lengths is sufficient 
to determine all distances, and thereby fix the linear scale of the 
Universe. This ``time delay'' phenomenon and its utility for the 
determination of $H_0$ was the subject of a recent Workshop at the 
Nuffield Radio Astronomy Laboratories, Jodrell Bank, part of the 
University of Manchester's Department of Physics and Astronomy.

\mvs
\section{The arithmetic of lensing}

Virtually everything one might want to calculate for a gravitational
lens can be computed from the expression for the light travel time for
a photon traveling from a source to an observer (Narayan and
Bartelmann 1996, Blandford and Kundi\'c 1996). As can be seen from
the equation in figure 2, there are two contributions 
to the light travel time: a geometric contribution, proportional to
the square of the difference in angular position between where the
source would have appeared and where the image does appear, and a
gravitational contribution, which results from the gravitational
potential altering the effective index of refraction.  For a ``thin'' 
lens, the gravitational effects can be lumped together as a 2D projected 
potential by integrating the 3D potential along the line of sight. 
A 2D plot of this light travel time is called the arrival time surface;
the example in figure 3 shows isochrones.

Fermat's principle tells us that images of the source will appear
wherever the light travel time is a minimum, a maximum or a
saddle point.  To find these stationary points one takes the
gradient of the light travel time with respect to position on the
sky and sets it equal to zero.  This gives a vector equation,
often called ``the lens equation'', for the positions of the images
on the plane of the sky.  There may be one, three, five or more 
solutions to this equation, giving an odd number of images. Applying 
the lens equation to the arrival time surface of figure 3 one can 
readily identify the positions of five images, marked as dots.
The cross is the position of the source in the absence of the lens.
The time delay between the images is just the difference in ``height''
of the arrival time surface at the image positions.

Taking the gradient of each Cartesian component of the lens equation
separately, one finds expressions for the partial derivatives of the
source position with respect to the image position.  Assembling these
terms (which include second derivatives of the projected potential) as
a matrix gives the ``back projection matrix'', telling what size source
corresponds to what size image.  More useful is its inverse,
``magnification matrix'', which tells us how a source is distorted.  An
important point here is that for a potential which is sharply peaked
at its centre, the back projection factor is large (because of the
large second derivatives of the potential), and the magnification 
small.  While in principle one should see odd numbers of images, in
practice the images which ought to appear near the centres of the
lensing galaxies (maxima in the light travel time) are so strongly
demagnified that they are rarely detected.  Hence virtually all the
lenses one sees are classified as doublets or quadruplets, in seeming
contradiction of the requirement that there be an odd number of
images. Figure 3 shows the shapes and sizes of the five images 
produced by our lens from a small circular source. Because lensing 
conserves surface brightness of the source (a consequence of 
Liouville's theorem), the apparent sizes of the images are directly 
proportional to their magnifications. Note that -- as for actually
observed galaxy lenses -- our lens demagnifies the central image so 
that it is barely detected. 

The light travel times are proportional to the distance from the 
observer to the lens and the distance from the observer to the source 
and are inversely proportional to the distance from the lens to the 
source.  The relative sizes of these quantities can be determined from
the redshifts of the lens and the source, but the overall scaling factor
depends upon the Hubble constant, the value for which continues to be
a matter of some debate.  If the difference in light travel times can
be measured, and if the gravitational potential can be modelled, one
can solve for the Hubble constant.

\mvs
\section{Practical hurdles on the way to $H_0$}

Some 18 years have passed since the discovery of the first
gravitationally lensed quasar by Walsh and collaborators
(Walsh {\it et al.} 1979; Walsh, 1989).  If
measuring distances to such systems were really so straightforward,
one might reasonably wonder why questions regarding the cosmological
distance scale haven't long since been resolved.  As one might have
guessed from the name of the workshop, not all lenses are equally 
useful -- hence the quest for the ``golden lens''.
Perhaps the most frustrating fact is that, despite their 
reputation, most quasars are surprisingly non-variable; there are
beautifully multiply-imaged quasars which fail to show appreciable
flux variations over the timescale expected for a time delay. Other
strongly lensed systems are so symmetrical that the time delays are both
very short and very sensitive to small errors in the measurement of
the relevant positions. Still other systems are quite promising
except for the fact that either the lensing galaxy or the lensed
quasar is too faint for a redshift measurement. 

There have been sociological hurdles as well. The scheduling of 50 or
100 appropriately spaced observations for a time-delay monitoring
program is not always easy, especially with changes in instrument
configurations and the traditional scheduling scheme at optical
observatories whereby a single program is given a contiguous
block of several nights. 

Even if the foregoing problems are overcome and a near-perfect lens
system is found with its properties carefully measured, there 
still remains a challenge of constructing a model for the gravitational 
potential of the system. The task is not trivial because the observables
provide a rather limited set of constraints for the model, leaving us
to rely on various assumptions about what a galaxy should look like
in general. In addition, it often happens that the main lensing galaxy 
is embedded in a group or cluster of galaxies, like in the case of 
Q0957+561. Such additional mass concentrations around the lens make model 
building even more difficult, and the resultant $H_0$ even more uncertain.

Because both observational and theoretical aspects of the problem are 
challenging and time-consuming there has been something of a 
chicken-and-egg problem, with observers reluctant to embark upon a 
monitoring campaign without first having a good model, and model 
builders reluctant to make detailed models before time delays have been 
measured.

One has to start somewhere, and the recent CLASS survey goes a long 
way towards addressing the observational challenges.

\mvs
\section{CLASS radio lens survey}

The past couple of decades have produced a number of large comprehensive
surveys in astronomy. All the ``early'' lenses were serendipitously 
discovered during some of these surveys; for example the original double 
quasar Q0957+561 A, B was discovered during the optical spectroscopic 
follow-up of Jodrell Bank 966 MHz sources.  The quadruple system 
Q2237+0305, 
another famous member of the lensing family, was discovered in the 
Center for Astrophysics (CfA) galaxy redshift survey. Only a handful 
of surveys have been specifically designed to uncover new lenses, or 
have finding new lenses as one of their main goals. The MIT-Green Bank 
5 GHz, Parkes-MIT-NRAO 5 GHz, and the Hubble Space Telescope (HST)
Snapshot surveys are among these. 

By far the largest of such surveys is the ongoing CLASS, an acronym
which stands for the Cosmic Lens All Sky Survey (Myers {\it et al.} 1995).
This effort is the follow-on to the Jodrell Bank VLA Astrometric Survey 
(JVAS, Patnaik 1993).  The CLASS collaborators, including astronomers 
from Jodrell Bank, the Netherlands and the US,
have obtained radio images of some 10 000 flat spectrum radio 
sources using the Very Large Array in Socorro, New Mexico. Only by 
heavily automating their data reduction were the investigators able 
to sift through the enormous volume of data gathered to identify the 
most likely candidates. A dozen or so of these have proven probable
lenses. The CLASS candidates have been followed up at higher angular
resolution using the MERLIN interferometric array, and at optical and
IR wavelengths using the William Herschel Telescope on La Palma and
the Keck Telescope on Mauna Kea.  Optical images of many of these
newly discovered lenses have been taken with the HST, most of them 
clearly showing the galaxy responsible for the lensing. 

New lenses from the CLASS and JVAS surveys have brought new surprises. 
Contrary to the theoretical wisdom of the past decade, most of the 
galaxy lenses in this sample are classified as spirals or S0s and not 
ellipticals. The two best examples of spiral galaxy lensing are the 
objects B0218+357 and B1600+434. The 
disk in the latter case appears to be almost exactly edge-on. Spurred 
by these observations, theorists have now turned to the problem of 
modelling spirals. Unlike elliptical galaxy models whose potential is 
well represented by some variation on the basic isothermal sphere, the 
spiral galaxies also have a flattened disk which has to be taken into 
account because the lensing cross section increases dramatically when 
disk is turned edge-on. 

It appears that Q0957+561, the first lens to be discovered, was not
unique in being a member of a group or cluster of galaxies. Several 
other lenses have companions in close projection to the main galaxy lens,
some with confirmed matching redshifts. In hindsight this is not 
surprising -- galaxies are known to cluster, and the larger optical depth 
presented by close collections of galaxies increases magnification,
thereby enhancing the chance that the background object is included in 
flux limited surveys. 

The CLASS has also uncovered a few unusual and puzzling lenses.
Perhaps the most spectacular of these is the new lens B1933+504, which 
shows a total of 10 radio images.  Observationally the images can be 
grouped into those of similar radio spectra and compactness. 
With the help of analytical models, 
one concludes that one is seeing three components of an active radio 
galaxy; the compact flat spectrum central core and one of the steep 
spectrum lobes are both quadruply imaged, while the second radio lobe 
is doubly imaged. The abundance of multiple images makes this system 
ideal for modellers; it fails to qualify as a ``golden lens''
because the lack of an optical counterpart, even with the
HST, means that the redshift of the source is unknown.

The JVAS lens B0218+357 is an excellent system for time delay 
measurements. It consists of two components which have been observed to 
vary in radio flux (VLA 15 GHz), the degree of polarization, and 
the polarization position angle. As the variations can be tracked by
all three methods, and are relatively large, 10\%, it should be
possible to measure the time delay to high fractional accuracy. 
The current estimate, $12\pm 3$ days, is likely to improve rapidly. 
The modelling of this system should be helped by the existence of a 
radio ring, and the fact that the lensing galaxy, a face-on spiral, is 
clearly visible. The galaxy contains neutral and molecular gas, seen in
absorption against the background radio images. However, before B0218+357 
is modelled in detail two aspects of the systems must be understood: the 
separation of the two radio images and two optical components are 
different by 10\% (significant given the size of the error-bars), 
while the flux ratio of the images differs by a factor of nearly 40 
between the optical and the radio. Dust, known to exist in 
spiral galaxies, may be in part responsible for these observations. 

JVAS lens B1422+231, a quadruple system with three very bright images 
and a fourth faint one, presents a serious problem for modellers.
The image positions can be easily fitted by a range of models, but
when radio flux ratios are taken into account there is no acceptable
analytical fit for the galaxy. It is generally thought that radio
flux ratios should provide useful additional constraints for 
the overall potential of the galaxy. This is because the radio-emitting 
region of the source quasar is substantially larger than the lensing
cross-sections of individual stars in the lensing galaxy, so that
microlensing by stars should not introduce flux variations in the 
images above and beyond those attributed to the smooth potential of 
the galaxy as a whole. Moreover, radio flux is not obscured by dust.
Thus it appears either the models tried so far are inadequate for 
B1422+231, or that its substructure significantly affects image 
magnifications.

CLASS B1608+656, shown in figure 4, is another potential ``golden lens''. 
The source is unusual because it is not a quasar but a post-starburst 
radio galaxy. The extended nature of the galaxy gives rise to a broken 
ring observed in the lensed image. The system is being monitored with 
the VLA, and the time delays are reported to be of the order of a few 
weeks. A potential problem is that the lensing galaxy does not seem 
simple: it is either an interacting system, or its image is split by 
an obscuring dust lane.

The present sample of CLASS lenses, though exciting and fascinating,
is not yet a complete sample in a statistical sense. This is because
the quadruplets and higher multiplicity systems are much easier to
confirm as lenses using geometry, without the help of optical
counterparts and redshifts.  Double systems, as well as larger 
separation lenses with several arcseconds between images, and lenses 
without obvious galaxies are more difficult to confirm as lenses.

As $H_0$ can be determined from an individual lens system, (or a 
collection of lenses to take into account modelling and observational 
errors) the current CLASS results can already be used for that purpose,
while some other cosmological applications (those that require 
statistically homogeneous samples) must wait until the CLASS survey
is complete.

\mvs
\section{Recent progress}

The progress reported at the Jodrell Bank meeting was along three
principal fronts. The first of these was the discovery of a large
number of new lenses, the most interesting of which are described above.

The second area in which considerable strides have been made has been
in the measurement of time delays.  The relatively small amplitude of
flux variations demands scrupulous attention to calibration in carrying
out observations.  By the end of the meeting the number of secure
measurements of time delays had doubled from two to four. The best
determined time delay is that for the oldest known system, Q0957+561.
Using two year's worth of data obtained in 1995-1996 at the Apache 
Point Observatory in New Mexico and a range of different methods 
to analyze the light curves (see figure 5), Kundi\'c, Wambsganss and 
collaborators (Kundi\'c {\it et al.} 1997) found the time delay between 
A and B to be $417\pm 3$ days, confirming the earlier, less robust 
result of Schild and Cholfin (1986). The time delays in a quadruple
PG1115+080, the second gravitational lens to be discovered, are 
currently determined to $\sim 10\%$ accuracy. This is likely to
improve with the addition of more data. The lenses B0218+357 and 
B1608+656 now have measured time delays, from radio observations, for 
which the fractional uncertainty is also likely to improve rapidly.

A third front on which progress was reported was in modelling the
gravitational potentials, which make a substantial contribution to the
total time delay.  The potentials are constrained by the observed
pattern of multiple images.  But there are few images, so the potentials 
are poorly constrained. The lens equation shows that the image positions 
depend upon derivatives of potential rather than on the potential itself.
But it is the potential that determines time delays; reconstructing it 
from a small number of derivatives is an ill-posed problem.

In a few cases some extra constraints to aid reconstruction are provided 
by the additional images observed in the system. For example, VLBI images
of Q0957+561 show a core and a radio jet with five blobs, each of which 
is also doubly imaged. These extra constraints are much needed in this 
system where the lens consists of a galaxy plus a parent cluster and 
hence requires more parameters for even minimal modelling. There is an 
intrinsic degeneracy in the
modelling because from the image data alone we do not know how the total 
mass causing the image separation is split up between the galaxy and the 
more smoothly distributed cluster mass. This ``mass-sheet degeneracy'' 
which translates into an uncertainty in $H_0$, can be broken in a number
of ways. One is by observing additional images, such as jet components in 
Q0957+561. Another, is by studying the lens itself: 
a measured velocity dispersion within the main galaxy can be, albeit 
with caveats, converted into its mass. Alternatively, a weak shear map 
of the background galaxies, or the velocity dispersion of galaxies within 
the cluster, can be used to estimate cluster mass. It is encouraging 
that different ways of breaking the degeneracy in Q0957+561 lens lead 
to nearly the same estimates of $H_0$. 

A further strategy to surmount the problem of limited constraints is 
through the modelling of the lensing mass. Some groups have adopted 
``reasonably'' parameterized galaxy models, while others constructed 
non-parametric models imposing additional constraints requiring 
smoothness and a degree of central concentration. The ultimate test 
is whether modellers using different techniques agree on the predicted 
time delays. Agreement appears to be best in the case of Q0957+561, 
which is also the lens with the best measured time delay.

The current status of lens time-delay measurements is given in table 1,
where we have listed separately the fractional uncertainties associated 
with the time-delay measurements and with the models for each of the 
four lenses with secure time delays. In the last column we show the 
derived value of the Hubble constant for three of the four systems.  
Agreement is quite good, but with the simplifying assumptions made in 
modelling, a mere three systems seems hardly adequate.  The onus for 
improving this state of affairs falls equally on those measuring delays 
and on those making models.  

\mvs
\section{Convergence of methods}

Gravitational lens time delays are hardly the only means of measuring
extragalactic distances and there has been considerable talk recently
(Kennicutt 1996) about the convergence of the many different methods.
The crucial rung on what is known as the ``cosmological distance ladder'' 
is the period-luminosity relation of Cepheid variable stars 
(Jacoby {\it et al.} 1992). Their use as distance indicators dates back to
the historic Curtis-Shapley debate of 1920 (Trimble 1995) where Cepheids 
were used, though somewhat erroneously, to argue 
{\it against} the extragalactic nature of ``spiral nebulae''. The modern 
distance ladder (really a coupled set of ladders) relies ultimately on 
parallax measurements, but includes as a key link the calibration of the 
period-luminosity relation for Cepheids. There is considerable debate 
at present about the extent to which this relation depends upon the 
abundances of elements heavier than helium, since these abundances can 
be quite different in other galaxies from the values measured locally 
in the Milky Way. Moreover the distances to Milky Way Cepheids have 
become more uncertain rather than less uncertain with the first published 
results from the Hipparcos astrometric satellite. 

Unlike the distance ladder, the gravitational lensing method measures 
distances on truly cosmological scales, and does so directly, 
bypassing all the rungs of the ladder together with their associated 
errors. One other method can claim that: the Sunyaev-Zel$^\prime$dovich 
technique basically compares the line of sight extent of the cluster, 
estimated from the properties of its hot X-ray emitting gas, to the 
cluster's $H_0$-dependent projected size. Systematic uncertainties 
associated with projection effects and gas clumpiness tend to 
underestimate $H_0$. 

While at present the gravitational lenses give distances which are no 
more accurate than the other distance indicators, they are at least 
subject to a completely different set of systematic errors. The major
recognized source of these is lens modelling, including the mass-sheet 
degeneracy. These systematic uncertainties have not yet been explored 
mostly because lens models used until recently were all rather similar. 
The new and better data from the CLASS survey and 
other campaigns have prompted a more comprehensive look at 
modelling, a trend which will soon allow us to assess critically the 
associated errors. The mass-sheet degeneracy can in principle be mostly 
resolved with independent measurements of the smooth cluster mass 
around the images. 

Given that all methods are subject their own systematic uncertainties,
it is essential to pursue all the available methods in order to reach
a convergence. 

\mvs
\section{Conclusions}

Several requirements need to be met in order to measure $H_0$ from
gravitational lensing. First, a multiply imaged lens system has to be 
found that possesses a number of desirable characteristics, like well 
resolved images, measurable source variability, easily visible lensing 
galaxy and optical image counterparts. Once a system is found, it must 
be extensively observed to get well sampled image light curves and
accurate astrometric information. Several of the new CLASS lenses and 
several previously known satisfy these requirements to 
various degrees; a few appear to be potential ``golden lenses''. 

Next, a good analytical model for the gravitational potential of the 
lensing galaxy/
\newline
galaxies is needed. Because the observed parameters 
of the lens are not nearly enough to constrain the model properly, many 
different approaches must be explored, so as to quantify carefully the 
uncertainties arising from modelling. Since several parametric and 
non-parametric modelling techniques have been published recently
a comparison of these will soon be possible. 

Finally, after 
random and systematic errors have been assessed, the comparison of model 
predicted and observed time-delays between images will yield the value 
of the Hubble constant. Taking stock of the recent observational and 
theoretical developments, the prognosis for the gravitational lens method
is excellent.

\vskip 0.2in

Originally published in volume 38:5 of Astronomy \& Geophysics, the 
Journal of the  Royal Astronomical Society, by the Institute of Physics 
Publishing Co.

\vskip 0.2in

\nid {\it
L. L. R. Williams is at the Institute of Astronomy, Cambridge, Britain,
and 
\newline
P. L. Schechter is at the Massachusetts Institute of Technology,
Cambridge, MA, USA. Proceedings of the Golden Lenses Workshop are on 
line at: 
\newline
http://multivac.jb.man.ac.uk:8000/ceres/workshop1/proceedings.html
\newline
~Also the CLASS Home page (Caltech version):
\newline
http://astro.caltech.edu/ $\tilde{\s}\!$cdf/class.html
\newline
~and a Gallery of lens images (Compiled by Prof. J. Surdej):
\newline
http://astra.astro.ulg.ac.be/grav\_lens/
\newline
~A new CfA-Harvard compilation of lens images and related information:
\newline
http://cfa-www.harvard.edu/glensdata/
}

\newpage

{
\begin{table}   
\begin{center}
\begin{tabular}{|c|c|c|c|}
\hline
      &             &  $\Delta t / t $     &             \\
Lens system & observed & predicted & best estimate of $H_0$ \\ \hline
Q0957+561   & 1\%  & 10\% & 61  \\
PG1115+080  & 10\% & 15\% & 53  \\
B0218+357   & 25\% & 30\% & 70  \\ 
B1830-211   & 20\% &  ?   & ?   \\ \hline
\end{tabular}
\caption[]{Observed and predicted uncertainties in the time delays
       between images in four gravitationally lensed systems
       with most secure measured time delays. The best estimate 
       for the $H_0$ is quoted in the first three cases.
}
\label{table1}
\end{center}
\end{table}
}

\begin{figure}
\vbox{
\centerline{
\psfig{figure=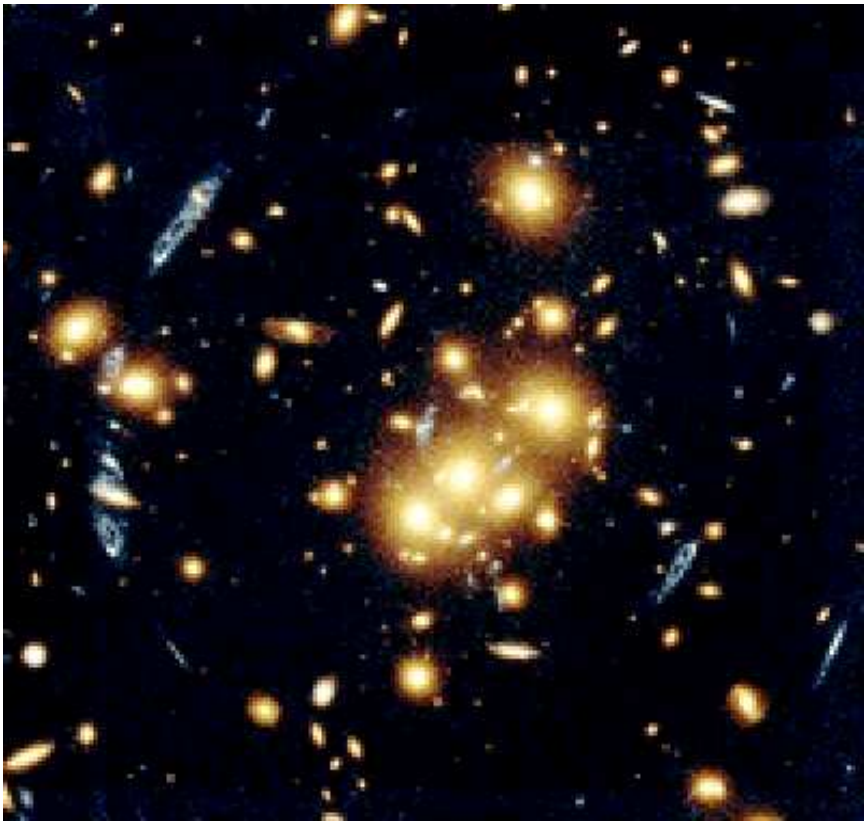,width=4.5in,angle=0
}
}
\caption[]{A cluster of galaxies CL0024+1654. A single background source,
a distant ring galaxy, is being strongly lensed by the cluster into
five images, which are readily recognized by their blue colour. The
cluster member galaxies appear reddish yellow. The cluster distorts 
each of the five images somewhat differently, but these distortions are 
easily understood as the effect of the cluster's gravitational potential.
Figure courtesy of Wes Colley (Princeton U.), Tony Tyson (Lucent 
Technologies), and Ed Turner (Princeton U.).
\label{Figure1}
}
}
\end{figure}

\begin{figure}
\vbox{
\centerline{
\psfig{figure=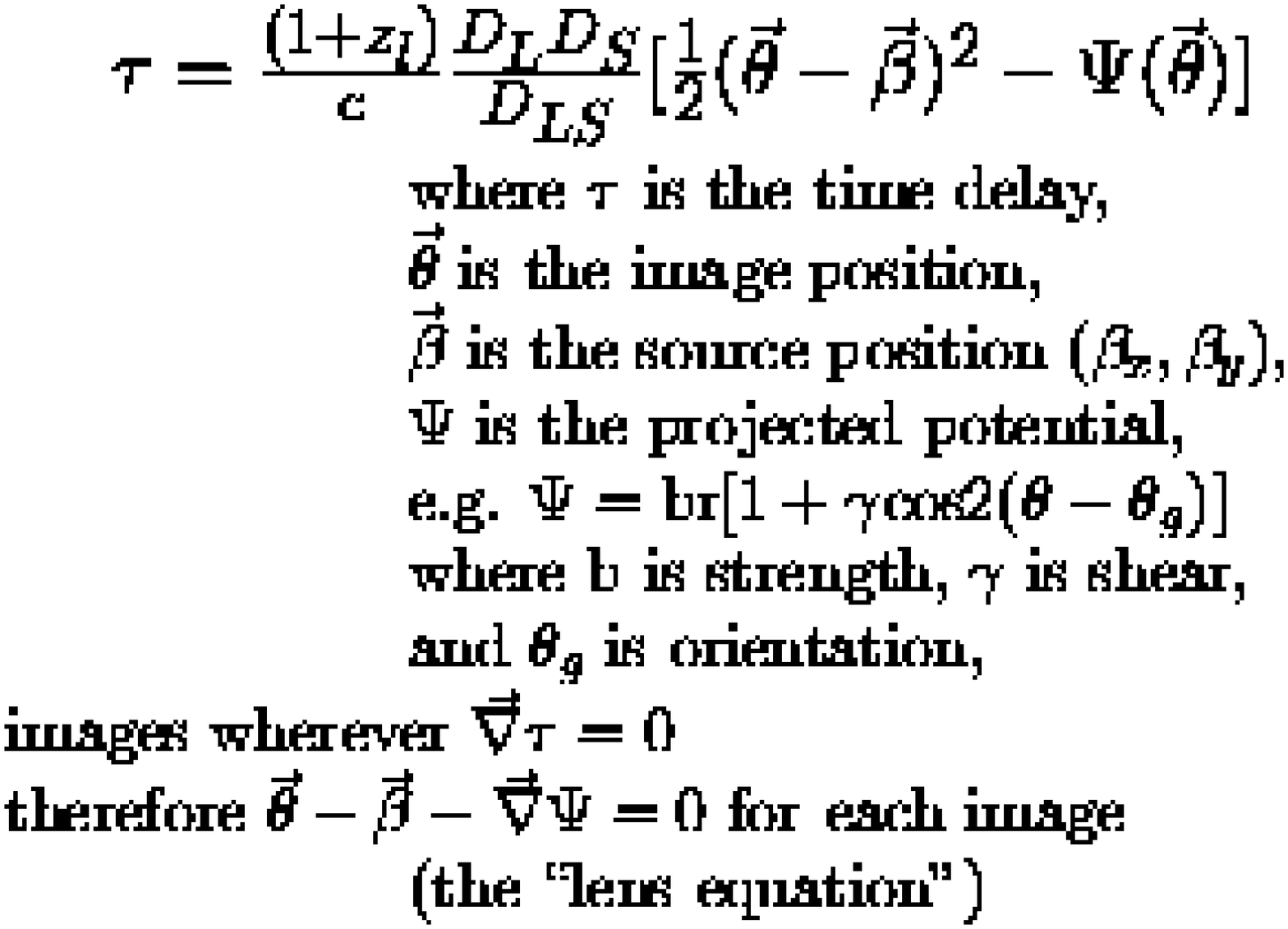,width=3.5in,angle=-90
}
}
\caption[]{The equation for light travel time for a photon traveling 
from a source to an observer. The two contributions to the light travel 
time, represented by the two terms in parentheses, are the geometrical 
and gravitational contributions respectively. 
\label{Figure2}
}
}
\end{figure}

\begin{figure}
\vbox{
\centerline{
\psfig{figure=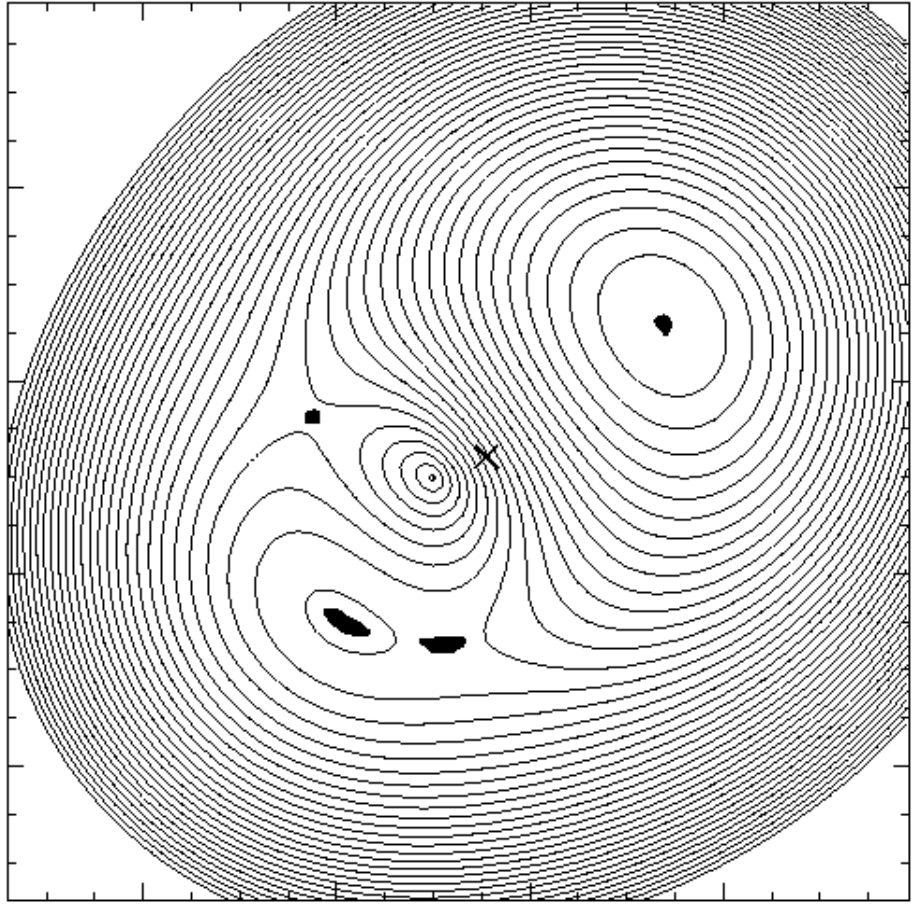,width=6.0in,angle=0
}
}
\caption[]{A contour plot of the light travel time, or the arrival time 
surface from equation of Figure 2. The images are formed at the maxima, 
minima and saddle points of this surface. There are five images here, 
marked by black dots. The sizes and shapes of these dots are proportional
to those of the images. The cross is the position of the source in the 
absence of the lens. The time delay between the images is the 
difference in ``height'' of the arrival time surface at the image 
positions.
\label{Figure3}
}
}
\end{figure}

\begin{figure}
\vbox{
\centerline{
\psfig{figure=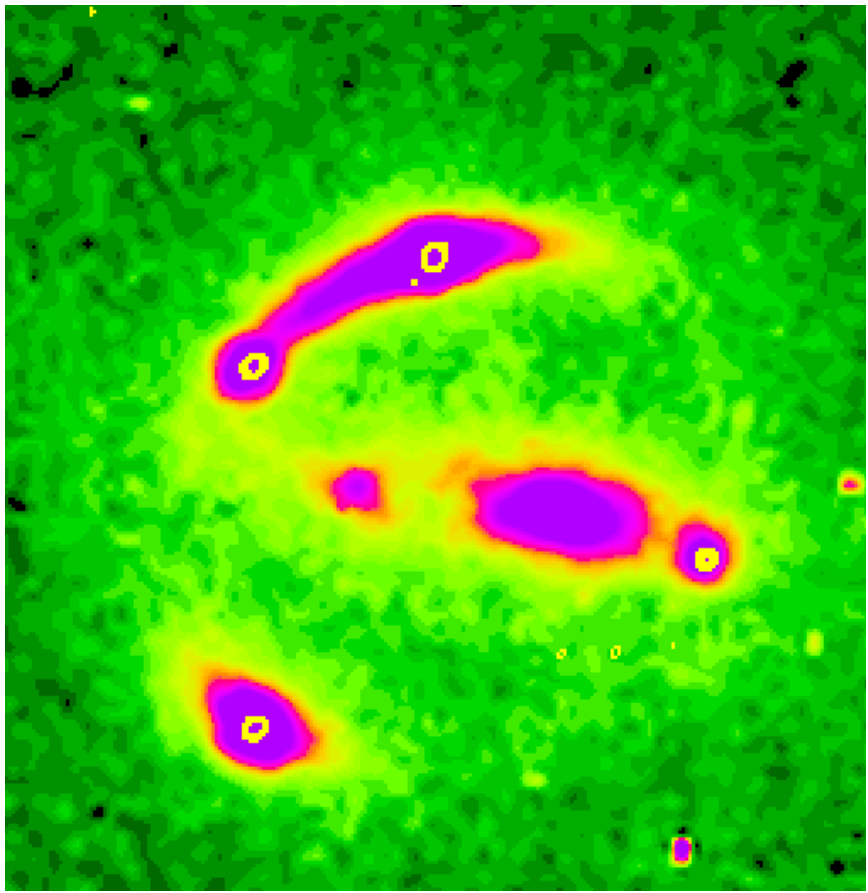,width=4.5in,angle=0
}
}
\caption[]{A gravitationally lensed QSO, B1608+656, from the CLASS 
survey. The source is a post-starburst radio galaxy with extended 
optical structure, and compact radio-emitting core. The source is split 
into four images, shown by the yellow contours of 5 GHz MERLIN radio map. 
The underlying image is the flux intensity map of the optical 814 
nanometre HST image. The central optical feature, with no radio 
counterpart, is the lensing galaxy. (Neal Jackson, Jodrell Bank, 
University of Manchester).
\label{Figure4}
}
}
\end{figure}

\begin{figure}
\vbox{
\centerline{
\psfig{figure=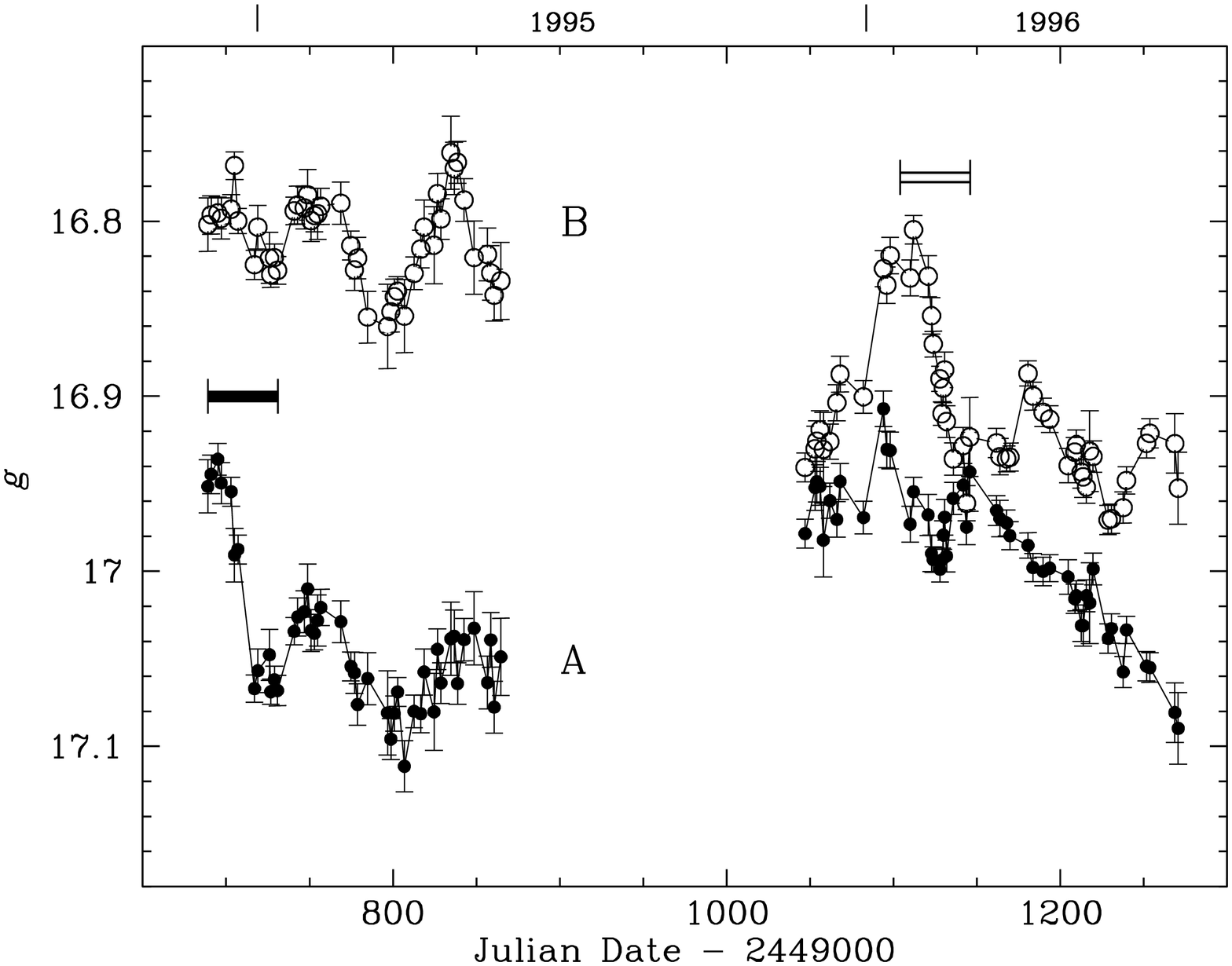,width=3.5in,angle=-90
}
\psfig{figure=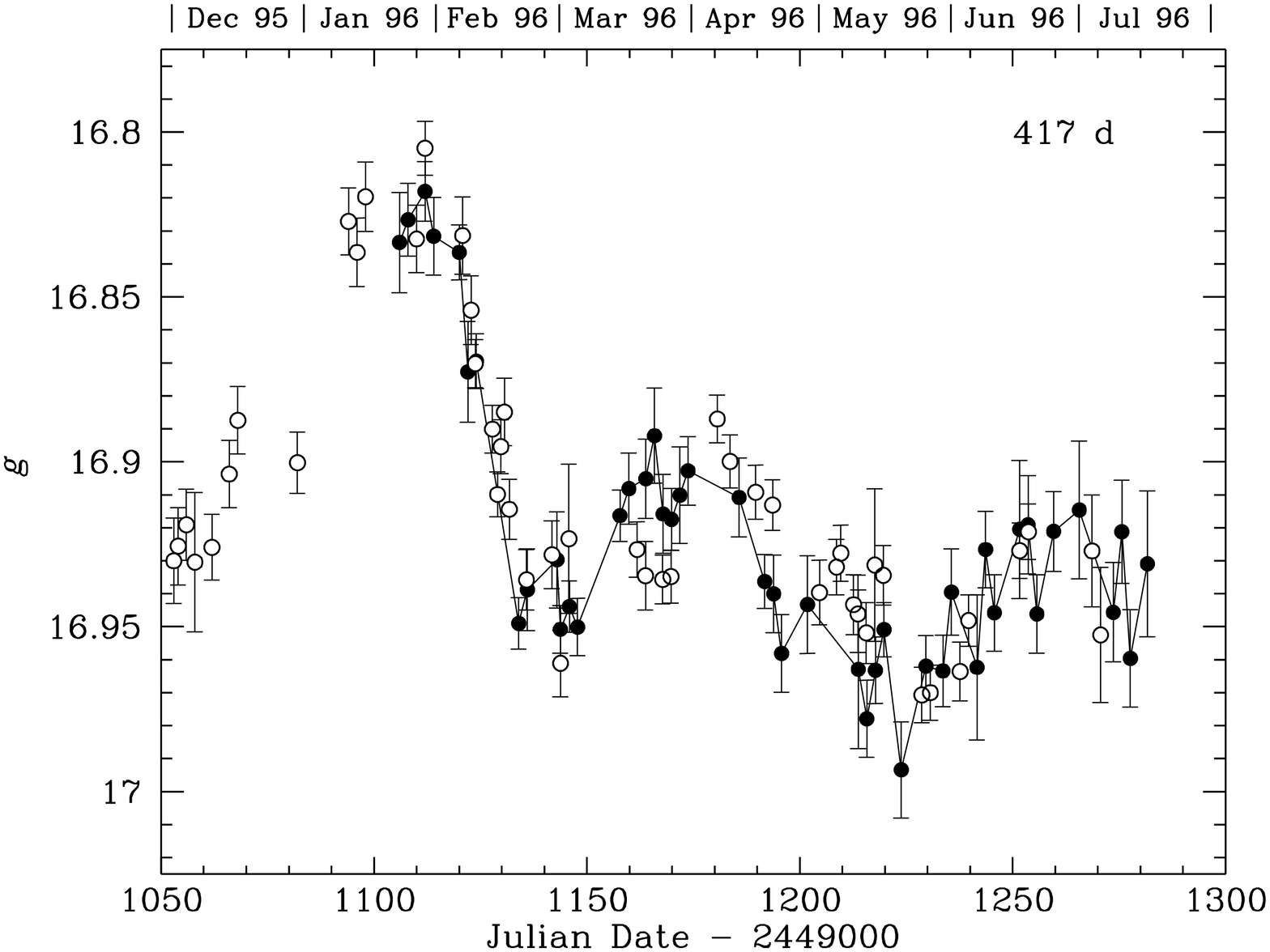,width=3.5in,angle=-90
}
}
\caption[]{Lightcurves of the two images of the gravitationally lensed 
quasar, Q0957+561; filled and empty circles represent images A and B 
respectively. The left panel shows the lightcurves as a function of 
Earth time (bottom axis: Julian date; top axis: year). The filled 
bar at the beginning of the 1995 season indicates when image A 
suddenly decreased in brightness by about 0.15 magnitudes (15\%). 
The empty bar at the beginning of the 1996 season shows that the 
decrease also occurred in image B, but 417 days later. This event,
together with other features in the lightcurves, allowed T. Kundi\'c
and collaborators to obtain the most precise measurement of time delay 
(accuracy of 1\%) of all known gravitationally lensed systems. The right
panel displays lightcurve of image A, advanced by the optimal value of 
the time delay, 417 days, and offset by -0.12 magnitudes, overlayed on 
the 1996 image B data. Figure courtesy of T. Kundi\'c (Caltech).
\label{Figure5}
}
}
\end{figure}


\begin{thebibliography}{99}

\bibitem {}
Blandford R., \& Kundi\'c, T. 1996, in
{\it The Extragalactic Distance Scale},
eds. M. Livio, M. Donahue, and N. Panagia, Cambridge University Press

\bibitem{}
Jacoby, G. H., {\it et al.} 1992, PASP, 104, 599

\bibitem{}
Kennicutt, R. 1996, Nature, 381, 555

\bibitem{}
Kundi\'c, T., {\it et al.} 1997, Ap, 482, 75

\bibitem{}
Myers, S. T., {\it et al.} 1995, ApJ, 447, L5

\bibitem{}
Narayan, R., \& Bartelmann, M. 1996, 
Lectures held at the 1995 Jerusalem Winter School;
also available at http://xxx.soton.ac.uk/ps/astro-ph/9606001

\bibitem{}
Patnaik, A. 1993, in
{\it Gravitational Lenses in the Universe},  eds. J. Surdej {\it et al.} 
(Li\`ege: Institut d'Astrophysique), p. 311

\bibitem{}
Schild, R. E. \& Cholfin, B. 1986, ApJ, 300, 209

\bibitem{}
Trimble, V. 1995, PASP, 107, 1133

\bibitem{}
Walsh, D., Carswell, R. F., \& Weymann, R. J. 1979, 
Nature, 279, 381

\bibitem{}
Walsh, D. 1989, in ``Gravitational Lenses", 
{\it Lecture Notes in Physics}, 330, eds.
J. M. Moran {\it et al.}
Springer-Verlag, p. 11

\end{thebibliography}
\end{document}